\documentclass{emulateapj}
\def\LCDM{$\Lambda$CDM~}
\def\zp{z_{phot}}
\def\bzp{\bar{z}_{phot}}
\def\geqsim{\lower.73ex\hbox{$\sim$}\llap{\raise.4ex\hbox{$>$}}$\,$}
\def\leqsim{\lower.73ex\hbox{$\sim$}\llap{\raise.4ex\hbox{$<$}}$\,$}

\def\Mpch {~h^{-1}~{\rm Mpc}}
\def\kpch {~h^{-1}~{\rm kpc}}

\newcommand{\myemail}{admyers@astro.uiuc.edu}
\newcommand{\Hii}{\ion{H}{2}}

\slugcomment{ApJ, in prep \today}
\shorttitle{Clustering of 300,000 Photo-Classified QSOs}
\shortauthors{Myers et al.}

\begin{document}

\title{Clustering Analyses of 300,000 Photometrically Classified Quasars--{\rm II}.  The Excess on Very Small Scales}

\author{Adam D. Myers\altaffilmark{1,2}, Robert J. Brunner\altaffilmark{1,2}, Gordon T. Richards\altaffilmark{3,4}, Robert C. Nichol\altaffilmark{5}, Donald P. Schneider\altaffilmark{6} and Neta A. Bahcall\altaffilmark{7}}

\email{\myemail}

\altaffiltext{1}{Department of Astronomy, University of Illinois at Urbana-Champaign, Urbana, IL 61801}
\altaffiltext{2}{National Center for Supercomputing Applications, Champaign, IL 61820}
\altaffiltext{3}{Department of Physics \& Astronomy, The Johns Hopkins University, 3400 N Charles St, Baltimore, MD 21218}
\altaffiltext{4}{Department of Physics, Drexel University, 3141 Chestnut Street, Philadelphia, PA 19104}
\altaffiltext{5}{ICG, Mercantile House, Hampshire Terrace, University of Portsmouth, Portsmouth, P01 2EG, UK}
\altaffiltext{6}{Department of Astronomy and Astrophysics, 525 Davey Laboratory, Pennsylvania State University, University Park, PA 16802}
\altaffiltext{7}{Department of Astrophysical Sciences, Princeton University, Princeton, NJ 08544}

\begin{abstract}
We study quasar clustering on small scales, modeling clustering amplitudes using halo-driven dark matter descriptions. From 91 pairs on scales $<35\kpch$, we detect only a slight excess in quasar clustering over our best-fit large-scale model. Integrated across all redshifts, the implied quasar bias is $b_Q = 4.21\pm0.98$ ($b_Q = 3.93\pm0.71$) at $\sim$18$\kpch$ ($\sim$28$\kpch$). Our best-fit (real-space) power index is $\sim$-2 (i.e., $\xi(r) \propto r^{-2}$), implying steeper halo profiles than currently found in simulations.  Alternatively, quasar binaries with separation $<35\kpch$ may trace merging galaxies, with typical dynamical merger times $t_d\sim(610\pm260)m^{-1/2}h^{-1}$Myr, for quasars of host halo mass $m\times10^{12}~h^{-1}M_\sun$.  We find UVX quasars at $\sim$28$\kpch$ cluster $>5$ times higher at $z > 2$, than at $z < 2$, at the $2.0\sigma$ level. However, as the space density of quasars declines as $z$ increases, an {\em excess} of quasar binaries (over expectation) at $z > 2$ could be consistent with {\em reduced} merger rates at $z > 2$ for the galaxies forming UVX quasars. Comparing our clustering at $\sim$28$\kpch$ to a $\xi(r)=(r/4.8\Mpch)^{-1.53}$ power-law, we find an upper limit on any excess of a factor of $4.3\pm1.3$, which, noting some caveats, differs from large excesses recently measured for binary quasars, at $2.2\sigma$. We speculate that binary quasar surveys that are biased to $z > 2$ may find inflated clustering excesses when compared to models fit at $z < 2$. We provide details of 111 photometrically classified quasar pairs with separations $<0.1'$. Spectroscopy of these pairs could significantly constrain quasar dynamics in merging galaxies.
\end{abstract}


\section{Introduction}
\label{sec:intro}
Galaxy formation and quasar activity are becoming irrevocably linked in the wake of mounting evidence that most local galaxies contain supermassive black holes \citep{Kor95,Ric98} with dynamical masses that correlate to many properties of the galaxies' spheroidal component (e.g., \citealt{Mag98,Sil98}). Galaxy mergers are a likely progenitor of new massive systems (e.g., \citealt{Hec86}), so, in turn, quasar formation is strongly linked to galaxy merger activity. At redshifts of $z~\geqsim~2$, where gas is abundant, the mergers of gas-rich galaxies provided ample baryons to fuel vigorous UVX quasar activity (e.g., \citealt{Hop06}).  As baryons were depleted in the universe the merging of gas-rich galaxies shifted to lower masses (e.g., \citealt{Cro05,Hop06}) and UVX activity became scarcer, consistent with the observed decline in UVX quasars since $z~\leqsim~2$ (e.g., \citealt{Cro04}).

Unified models of quasar and galaxy formation are physically complex, and require new constraints from  large datasets. In particular, as quasar formation is driven by mergers of gas-rich galaxies on dark matter halo scales, small-scale quasar clustering is almost certainly a tracer of galaxy merger activity (e.g., \citealt{Djo91,Sch93,Koc99,Mor99}). It is not yet clear whether quasar activity is triggered in both of two merging galaxies or, rather, if quasar pairs are just tracing highly biased regions of the universe where many different merger events have occurred within a small volume. Of course, these two processes might dominate on different scales (e.g., \citealt{Hop06}). In either case, models of galaxy mergers and quasar formation should be usefully constrained by measurements of quasar clustering on small scales, and by any redshift evolution in quasar clustering on small scales. 

Measuring quasar clustering on small scales (where the total volume occupied by quasars dwindles) is limited by the number of known quasars. Further, in redshift surveys that utilize multi-object spectrographs, fiber collisions can restrict the number of pairs observed, even on arcminute scales (e.g., see, Figure~11 of \citealt{Cro04}). One method to improve clustering statistics on kiloparsec scales is to compile selections of quasar pairs observed for different purposes. \citeauthor{Hen06} (\citeyear{Hen06}; henceforth Hen06) used such an approach, and found that quasars cluster $\sim$10 times higher than expected on scales of $40\kpch$, growing to $\sim$30 times higher at $\sim$10$\kpch$. However, as the sample used by Hen06 was compiled via several techniques, and nominally, to target both gravitational lenses and quasars that trace the Ly$\alpha$ forest, their selection function is quite complex.

Many of the problems with studying quasar clustering on small scales could be circumvented by using a complete sample of photometrically selected objects that carried a high likelihood of being quasars. Until recently, star-quasar separation was insufficient to create such a sample (e.g.,$\sim$60\% efficiency in the 2dF QSO Redshift Survey; \citealt{Cro04}; henceforth 2QZ). However, the photometrically classified sample of \citet{Ric04} has highly efficient star-quasar separation ($\sim$95\%; \citealt{Ric04,Mye06,Mye07}), and offers a unique opportunity to study quasar clustering on kiloparsec scales using a sample with well-controlled statistical selection.

\citet{Mye06} demonstrated a proof-of-concept clustering analysis of $\sim$80,000 photometrically classified quasars at redshifts below $\sim$2.5.  In this series of papers, we extend this work to $\sim$300,000 such objects drawn from the fourth data release (DR4) of the Sloan Digital Sky Survey (SDSS; e.g., \citealt{Sto02,Aba03,Aba04,Aba05}).  Our goal in this paper is to quantify the amplitude of the small-scale clustering of quasars.  In a companion paper (\citealt{Mye07}; henceforth Paper1), we study the redshift and luminosity evolution of quasar clustering on larger scales ($\sim$50$\kpch$ to $\sim$20$\Mpch$).

The ``KDE'' data sample used in this paper, which is constructed using the Kernel Density Estimation (KDE) technique of \citet{Ric04}, is detailed in Paper1. The KDE technique is made viable by many technical aspects unique to the SDSS (e.g., \citealt{Yor00}), including superior photometry (e.g., \citealt{Fuk96,Gun98,Lup99,Hog01,Smi02,Ive04}), astrometry (e.g., \citealt{Pie03}), and data acquisition (e.g., \citealt{Gun06,Tuc06}). Our analysis methodology is discussed in depth in the Appendixes of Paper1. The only additional technique in this paper is our deprojection of angular power-laws of the form $\omega_{QQ}(\theta) = A\theta^{-\delta}$; we use the fact that $\gamma$, the slope of the real-space correlation function ($\xi(r)\propto r^{-\gamma}$) can be derived via $\gamma=1+\delta$ (see, e.g., \citealt{Mye06}). Throughout this paper, we use the \citet{Sch98} maps to estimate Galactic absorption. We adopt \LCDM with ($\Omega_m, \Omega_\Lambda, \sigma_8,\Gamma$, $h\equiv H_0/100{\rm km~s^{-1}~Mpc^{-1}}$)$ = (0.3,0.7,0.9,0.21,0.7)$ as our cosmology, where $\Gamma$ is the shape of the matter power spectrum. 

\section{Visual inspection of the Small-scale KDE pairs}
\label{sec:visinspec}

We do not explicitly mask the KDE sample for the standard mask holes available in the SDSS Catalog Archive Server, because the density of KDE objects is too low for significant numbers to fall in the masks. We have previously checked that this makes no difference to our analysis on large scales but it is rigorous to check whether our pairs are particularly biased on small scales, perhaps by fractured objects near bright stars. In pursuit of this, we visually inspected images of each of the 135 KDE pairs that have separations of $\Delta\theta<0.1'$. We find that only 1 of the 110 ($2.4''<\Delta\theta<6''$, $A_g < 0.21$) small-scale pairs in our analyzed area lies within a standard SDSS imaging mask; this object alone would have a negligible affect on our analysis or conclusions.

Visually inspecting the small-scale pairs led to the discovery of a number of dubious pairs that do not lie within standard SDSS masks but are obviously not genuine quasars. These are generally \Hii{} regions in various galaxies (a known contaminant of SDSS objects with star-like photometry). In all, we determined that 24 of the 135 $\Delta\theta<0.1'$ KDE pairs, and 18 of the 110 ($2.4''<\Delta\theta<6''$, $A_g < 0.21$) pairs in our analyzed area are not quasars. In general these dubious pairs have discrepant photometric redshifts, particularly compared to the redshift bins we use in this paper (in fact only three of these pairs lie in a shared bin, for bins as plotted in Figure~\ref{fig:biassmallscale}; all three in the $0.4\leq z_{phot} < 1.0$ bin). Throughout this paper we additionally correct our results to account for these objects that are really \Hii{} regions (as well as the one small-scale pair that lies in a standard SDSS mask hole), on our $\Delta\theta<0.1'$ scales of interest. The 111 KDE pairs with separations of $\Delta\theta<0.1'$ that are neither \Hii{} regions nor lie in a standard SDSS mask hole are recorded in Table~\ref{table:pairs}.

To check that masked objects and \Hii{} regions are not a general contaminant in the KDE data, we visually inspected 1000 randomly selected DR4 KDE objects. Only 4/1000 objects lay in masks or were \Hii{} regions. A contamination of $0.40\pm^{0.32}_{0.19}$\% is far lower than our expected stellar contamination, well within our error bars on all scales, and far too small to affect our analyses (other than as discussed on very small scales where \Hii{} regions can mimic quasar pairs).

\begin{figure}
\plotone{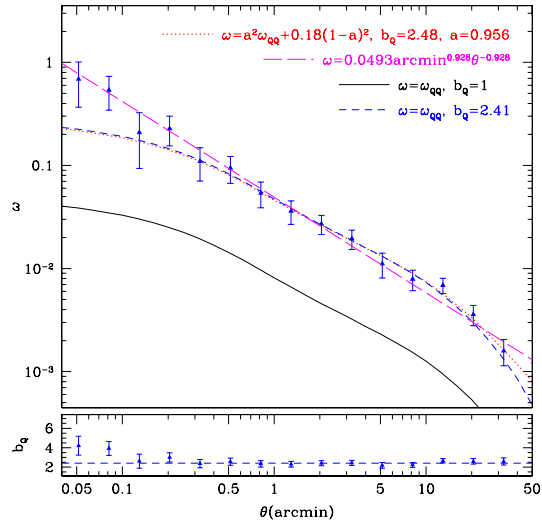}
\caption{The autocorrelation of all 299,276 ($A_g < 0.21$) DR4 KDE objects in our working area (see also Paper1). The short-dashed line is the best-fit bias model (Equation~B2 of Paper1) over the range $0.16'$ to $63'$ ($\sim$55$\kpch$ to $22\Mpch$ at $\bar{z}\sim$1.4). The dotted line is the best-fit bias model (over the same range) when including stellar contamination (Equations~B2 and C1 of Paper1, with $\omega_{SS}=0.18$). Note that including stellar contamination affects bias estimates less than the 3\% precision in the models (Smi03) used to derive $b_Q$. The solid line shows the expected clustering for a linear bias model, and the lower panel shows the bias relative to this model.  The long-dashed line is the best-fit power-law model over $0.16'$ to $63'$. Errors are jackknifed and fits include covariance.
\label{fig:allKDE}}
\end{figure}

\section{Excess Quasar Clustering on Small Scales}
\label{sec:smallscale}

In Figure~\ref{fig:allKDE} we show the autocorrelation function of objects in the DR4 KDE catalog. In combination, the points plotted in Figure~\ref{fig:allKDE} on scales $<0.1'$ rule out our best-fit large-scale bias model (Equation~B2 of Paper1, with $b_Q=2.41$) at $2.2\sigma$ (i.e., $>97$\%), using Poisson errors, which are valid on these scales (e.g., \citealt{Mye03,Mye05}). The innermost two bins plotted are, in fact, in better agreement with the (less physically motivated) best-fit large-scale power-law model ($\omega = 0.0493\pm0.0064\theta^{-0.928\pm0.055}$; $P_{<\chi^2}=0.483$). If the innermost bins are included in such a power-law fit, the best-fit model is barely changed ($\omega = 0.0469\pm0.0048\theta^{-0.912\pm0.045}$; $P_{<\chi^2}=0.622$).

We will proceed by measuring the amplitude of the innermost two points plotted in Figure~\ref{fig:allKDE} relative to our best-fit bias model and relative to a linear bias model. Of course, we don't necessarily expect the Smi03 models to be accurate on kiloparsec scales, but they nevertheless provide a reasonable phenomenological description of dark matter, and other authors may well combine such models with astrophysical factors to try and reproduce kiloparsec-scale quasar clustering. 

The innermost two bins ($\theta <0.1'$) in Figure~\ref{fig:allKDE} cover $2.4''<\theta<3.8''$ ($\sim$14 to $22\kpch$ at $\bar{z}=1.4$, the mean redshift of our sample) and $3.8''<\theta<6''$ ($\sim$22 to $35\kpch$). Contamination by pairs that are a single lensed source is unlikely, particularly for $\theta > 3''$ (e.g., Hen06). On these two scales, $\omega$ is $3.05\pm1.42$ and $2.65\pm0.96$ times higher than our best-fit large-scale bias model, and $17.7\pm8.2$ and $15.4\pm5.5$ times higher than a linear bias model. This suggests that adequate modeling of quasar clustering, at $\bar{z}\sim1.4$, will require $b_Q = 4.21\pm0.98$ ($b_Q = 3.93\pm0.71$) on $18\kpch$ ($28\kpch$) scales relative to the \LCDM description of Smi03. We find it is immaterial on these scales whether we include stellar contamination (see Appendix B of Paper1) in our modeling, or vary the large-scale cutoff of our fits, as such changes only affect our $b_Q$ estimates by a few percent---this is illustrated in Figure~\ref{fig:allKDE}, where the amplitude of the model autocorrelation on small scales is clearly unchanged by incorporating stellar contamination.

\begin{figure}
\plotone{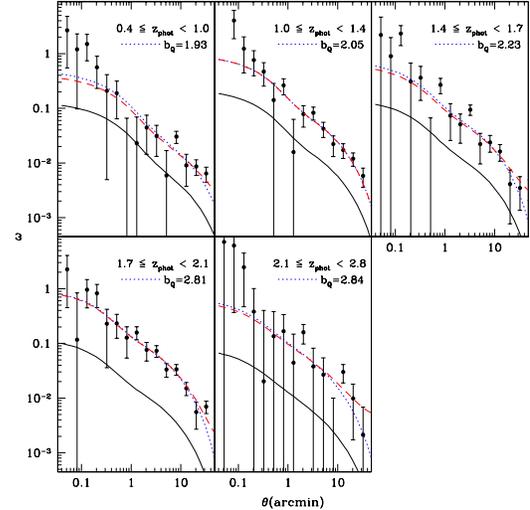}
\caption{Quasar clustering evolution with photometric redshift (see also Paper1). There are $\sim$65,000 quasars in each bin except for $2.1 \leq z_{phot} < 2.8$, which contains $\sim$28,000 quasars. The solid line is the expected \LCDM clustering derived from Smi03, with linear bias. The dotted line is our best-fit large-scale model where only the quasar bias, $b_Q$, is varied. The dashed line also incorporates stellar contamination. Errors in this plot are jackknifed and fits are made over scales of $0.16'$ to $63'$ including covariance. A scale of $0.16'$ to $63'$ is $\sim$55$\kpch$ to $22\Mpch$ in all bins except the lowest redshift bin ($\sim$50$\kpch$ to $20\Mpch$).
\label{fig:zbins}}
\end{figure}

\begin{figure}
\plotone{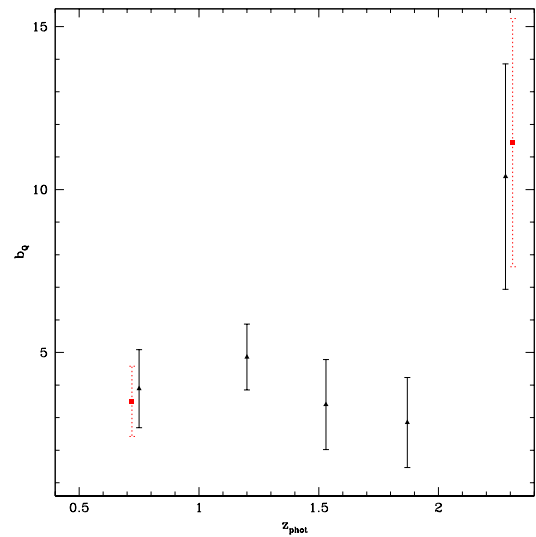}
\caption{The evolution of small-scale quasar clustering (see also Table~\ref{table:2}) integrated over scales of $2.4''<\Delta\theta<6''$ ($\sim$14 to $35\kpch$). The triangular points represent the quasar bias, $b_Q$ at a given photometric redshift, relative to the models of Smi03. Due to the photometric nature of the redshifts we use, estimates of $b_Q$ at the lowest and highest redshift are {\it at least} 10\% too high and 10\% too low, respectively (see section~3.2 of Paper1). We illustrate this 10\% correction with square points and dashed error bars. Points at other redshifts should not be biased at all by the use of photometric redshifts. Errors in this plot are based on Poisson estimates, which are valid on small scales (see, e.g., \citealt{Mye05}). \label{fig:biassmallscale}}
\end{figure}

\section{Evolution in Quasar Clustering on Small Scales}
\label{sec:smallevo}

In Figure~\ref{fig:zbins} we split our sample by photometric redshift and recalculate the quasar autocorrelation in each bin (see \citealt{Mye06} for validation of this technique). Figure~\ref{fig:zbins} suggests that excess quasar clustering is apparent on scales of $\leqsim~35\kpch$ at all redshifts. Again we apply the Smi03 models as a phenomenological description of the underlying dark matter. We allow the quasar bias $b_Q$ to evolve with redshift to best fit the data, and derive the implied $b_Q$ values on small scales (see Figure~\ref{fig:biassmallscale}).

Figure~\ref{fig:biassmallscale} (see also Table~\ref{table:2}) implies $\sim$5 times as many UVX quasar pairs existed at $z > 2$ than at $z < 2$, on scales of $\sim$28$\kpch$. This rises to $\sim$6.5 times as many relative to the large-scale expectations at each redshift. A model with constant $b_Q$ across $0 < \zp < 2.5$ is rejected by the data for the best fit $b_Q=4.14$ ($P_{<\chi^2}=0.083$). Although, $\chi^2$ may be misleading for so few constraints, our single data point at $z > 2$ rules out $b_Q=4.14$ at the $1.8\sigma$ level, consistent with the $\chi^2$ rejection. In Section~3.2 of Paper1, we noted that, because of our use of photometric redshifts, our points in the lowest and highest redshift bins are {\it at least} 10\% too high and 10\% too low, respectively (the other redshift bins are not biased at all by our use of photometric redshifts).  Incorporating this 10\% weighting into our analysis rules out the best, constant bias model of $b_Q=4.00$ at $\sim$95\%. The highest redshift ($z =2.28$) point in our weighted analysis rules out $b_Q=4.00$ at the $2.0\sigma$ level. If we combine our data at $z < 2$ into a single inverse-variance-weighted estimate ($b_Q = 3.82\pm0.59$) and determine whether this estimate would be rejected by the data point at $z > 2$, we again find that the level of rejection is $2.0\sigma$ (i.e., $95.2$\%). 

\section{Discussion}

\subsection{Overview and Comparison to Recent Studies}

For a UVX quasar sample with $\bar{z}=1.4$ we find the angular quasar autocorrelation is $\sim$3.0 ($\sim$2.7) times higher than our best-fit large-scale bias model and $\sim$17.7 ($\sim$15.4) times higher than a linear bias model at $\sim$18$\kpch$ ($\sim$28$\kpch$). Relative to the description of Smi03 (which, admittedly, may not be valid on the scales we are discussing), modeling of quasar clustering requires $b_Q = 4.21\pm0.98$ ($b_Q = 3.93\pm0.71$) at $\sim$18$\kpch$ ($\sim$28$\kpch$). On scales of $<35\kpch$, our best-fit bias model on larger scales is ruled out at $2.2\sigma$ (i.e., $>97$\%). Quasar bias therefore appears to be scale-dependent on scales $<35\kpch$.

Very few studies have been able to probe quasar clustering down to scales of a few arcseconds. The most impressive is Hen06, who detected excess quasar clustering of a factor of $\sim$10 ($\sim$30) at $\leqsim~40\kpch$ ($\leqsim~15\kpch$).  However, their sample was initially constructed from a candidate sample of gravitational lenses, complicating their color selection. Further, \citet{Hen06} measure their excess over a power-law derived from a UVX sample (PMN04) limited to $M_{b_J} < -22.5$ and $0.8 < z < 2.1$, quite different from their own color, magnitude and redshift selection. Notably, quasar clustering, grows significantly at $z > 2$ \citep{Cro05,Mye06}, and, although Hen06 attempt to correct for redshift evolution, they only do so to reflect PMN04 results at $z < 2.1$. Further, the Hen06 sample, in targeting tracers of the Ly$\alpha$ forest, may be biased toward $z > 2$. By contrast, the selection of our KDE sample is very uniform.  Our pairs will contain some stars, and our result is angular (as opposed to Hen06, who use redshift information).  However, as noted in section~\ref{sec:smallscale}, stellar contamination has little affect at $<1'$ and our sample size will offset the angular nature of our analysis, particularly as velocity effects can complicate redshift-space clustering measurements on kiloparsec scales.

Table~\ref{table:pairs} illustrates that 19 of the small-scale binary quasars detected by Hen06 are also included in our analysis. In fact, for objects with separations of $3''<\Delta\theta<6''$, all but 5 of the quasar pairs studied by Hen06 make our sample\footnote{The machine-readable tables in Hen06 suggest we miss 7 pairs, but 2 of these appear in multiple tables}. A sixth pair we miss, SDSSJ1034+0701, appears in Table~2 of Hen06 but is missing from the machine-readable data, and one member of this pair is at $g > 21$ so it could not make our sample. Of the 5 quasar pairs we miss; 3 have a member with a redshift of $z\geq2.4$ where the KDE selection is poor (only one of these is a true binary); 1 (a projected pair) has a member with redshift $z\sim2.2$, at the edge of our UVX color selection; and 1 is a quasar binary at $z\sim1.55$, only 1 member of which appears in the KDE catalog. As we find 98 likely quasar pairs with separations $3''<\Delta\theta<6''$, it is tempting to state that we find even more of an excess of quasar clustering on small scales than the large excess found by Hen06.  However, $\sim$100 quasar pairs on these scales is close to our expectation from large scales, and in section~\ref{sec:ultim}, we will explicitly demonstrate that our results are somewhat at odds with those of Hen06. 

\subsection{Implications for the Cores of Dark Matter Halos}

Alternative descriptions of dark matter on scales of $<35\kpch$ might explain small-scale UVX quasar clustering.  We find that a power-law fit to our data with slope $\delta = 0.912\pm0.045$ is acceptable ($P_{<\chi^2}=0.483$) on scales from $\sim$22$\Mpch$ down to $<35\kpch$, implying a real-space correlation function slope of $-(1+\delta)\sim-2$ (i.e., $\xi(r) \sim r^{-2}$; see, e.g., \citealt{Mye06}). This is significantly steeper on these scales than either an NFW (\citealt{Nav97}; $\xi(r) \sim r^{-1}$) or Moore (\citealt{Moo98}; $\xi(r) \sim r^{-1.5}$) halo profile. Smi03 include a halo term (e.g., \citealt{Pea00,Sel00}) within the models we have utilized, so a small-scale steepening in simulated halo density profiles could help explain quasar clustering at $<35\kpch$. \citet{Mas06} come to a similar conclusion in an analysis of the clustering of LRGs, also finding an excess over halo expectations at around $<35\kpch$, and a small-scale power-law consistent with $\xi(r) \sim r^{-2}$ on these scales.  

\subsection{The Merger Rate of Gas-rich Galaxies}

If the Smi03 description {\it is} valid at small scales, then UVX quasars are distributed more densely than dark matter at $<35\kpch$. If quasars formed in mergers between two gas-rich galaxies and, in certain stages of the merger event, black holes at the center of both galaxies can ignite (e.g., Figure~2 of \citealt{Hop06}), then UVX quasars should be highly biased on merger scales ($<35\kpch$ is not an unreasonable scale for the separation of merging galaxies).  In such a picture, quasars have two possible and entirely distinct UVX phases, one prior to the galactic merger (a UVX quasar binary) and one post-merger (a lone UVX quasar). Under this interpretation, the number of small-scale quasar binaries might be expected to diminish with cosmic time, particularly at $z < 2$ as gas in the universe is depleted and mergers shift to objects with insufficient mass to produce a UVX quasar visible (e.g., \citealt{Hop06}).  Such a decline at $z < 2$ is hinted at in Figure~\ref{fig:biassmallscale}. One might argue that, for UVX binaries on small scales, $b_Q$ {\it measures the number of merging gas-rich galaxies in the universe}. We will now assess this, admittedly qualitative, argument.

Figure~\ref{fig:allKDE} suggests a discrepancy at $\leqsim~35\kpch$ between dark matter models and quasars. We will proceed by inferring that at $>35\kpch$ we trace lone UVX quasars (shining after the merger of the galaxies that triggered the quasar) but at $<35\kpch$ we trace clustering enhancement due to the ensemble of UVX quasar binaries shining in both of two gas-rich galaxies prior to a merger. To estimate the merger timescale (the total time that quasars could be observed as a UVX pair pre-merger) we assume that the halos harboring UVX quasars merge on the order of a dynamical time

\begin{equation}
t_d \sim 2\pi\sqrt\frac{d_{Q}^3}{GM} 
\end{equation}

\noindent where $d_{Q}$ is the separation at which merging galaxies can begin to exacerbate UVX quasar behavior, $G$ is the gravitational constant and $M$ is the total mass associated with the merging galaxies. For the purposes of this discussion, we do not care whether both UVX quasars are turned on by the merger, one quasar is extant and triggers the other, or both quasars are extant; only that all quasars with separations $\leqsim d_{Q}$ trace galaxies that eventually merge. In essence, as soon as a quasar is within $d_{Q}$ of another, we cease to think of that object, whatever its history, as a lone quasar and treat it as a binary tracing a merger event.

The mass of the halo harboring a quasar post-merger will be the same as the total mass of the two parent systems. We will assume that pre-merger UVX pairs end up shining as a typical UVX quasar post-merger, and adopt a redshift-independent mass for UVX quasars' host halos of $M_{DMH} = m\times10^{12}~h^{-1}M_\sun$ where $m\sim3$--4 is consistent with recent measurements (e.g., \citealt{Cro05}; Paper1). Also adopting $d_{Q}=35\pm8\kpch$ from Figure~\ref{fig:allKDE} (with the uncertainty set by the bin width) we find $t_d\sim(610\pm260)m^{-1/2}h^{-1}$Myr. Given believable values of $m=3$ and $h=0.7$, this toy model is in very reasonable agreement with simulated merger timescales (e.g., Figure~2 of \citealt{Hop06}, Equation~10 of \citealt{Con06}).

To derive illustrative merger rates for the sample of UVX binaries we trace, we assume that all quasars at $d~\leqsim~d_{Q}$ are in galaxies that eventually merge. The implied comoving volume merger rate is then $\dot{\phi}_f = f_{m}t_{d}n_{Q}$ (e.g., \citealt{Con06}) where $f_{m}$ is the fraction of mergers being traced by quasars (i.e., {\em half} of the fraction of UVX quasars in pairs with $r~\leqsim~d_{Q}$) and $n_{Q}$ is the space density of quasars. We derive $n_{Q}$ simply from the number of quasars in a redshift bin divided by the comoving volume in that bin. However, we also compare with using the luminosity function of \citeauthor{Ric05} (\citeyear{Ric05}; only strictly valid for $M_g<-22.5$ quasars over the redshift range $0.4 < z < 2.1$), to derive the space density (see, e.g., equations 1, 2 and 3 of PMN04), denoting this $n_\phi$. As we do not have spectra, some of our quasar pairs will be chance alignments; and, further, as mergers are unlikely to occur after one dynamical time, $\dot{\phi}_f$ really represents an upper limit on merger rates.

To counteract the effect of chance alignments contaminating our quasar pairs, we can estimate merger rates via the real-space quasar autocorrelation (c.f., e.g., \citealt{Mas06}\footnote{Note that Equation 12 of the ApJ version of \citet{Mas06} should read $\Gamma_{LRG}n_{LRG}\sim0.6\times10^3$.}). The average number of UVX quasars within $\leqsim~d_{Q}$ of a companion is

\begin{equation}
N = 4\pi n_{Q}\int_{0}^{d_{Q}}r^2\xi_{QQ}(r){\rm d}r
\label{eqn:Nfrac}
\end{equation}

\noindent The real-space correlation function, $\xi_{QQ}(r)$, can be derived from the quasar bias we measured (sections~\ref{sec:smallscale}~and~\ref{sec:smallevo}) relative to Smi03 models (see also the Appendix of Paper1 and Smi03)

\begin{eqnarray}
\xi_{QQ}(r)=b_Q^2(z_{min},z_{max},<d_{Q})&&\int^{z=z_{max}}_{z=z_{min}}\left(\frac{{\rm d}N}{{\rm d}z}\right) \\ 
\times\int^{k=\infty}_{k=0}&&\Delta_{\rm NL}^2(k,z)\frac{\sin kr}{kr}\frac{{\rm d}k}{k}{\rm d}z \nonumber
\label{eqn:clar}
\end{eqnarray}

\noindent where we average $b_Q$ values over our redshift bins and out to $d_{Q}$, for (normalized) ${\rm d}N/{\rm d}z$ as in Paper1. It should be noted that although the Smi03 models may not be valid on our scales of interest, they are merely being used here to guide the redshift evolution of clustering amplitudes that have been established by the data. The comoving volume merger rate of the galaxies traced by UVX quasars is then

\begin{equation}
\dot{\phi}_{\xi} \sim \frac{Nn_{Q}}{t_{d}} \sim \frac{1}{610}Nn_{Q}m^{1/2}h^4{\rm Myr}^{-1}{\rm Mpc}^{-3}
\end{equation}

Values of interest arising from these models are displayed in Table~\ref{table:2}. In general, we consider $d_{Q}$ fixed at $35\kpch$, as changes scale systematically with $d_{Q}$. The relative values of $N/n_{Q}$ as a function of redshift suggest a strong increase in the number of quasar pairs at $z > 2$.  However the fraction of quasars in pairs that are tracing merging galaxies ($2f_{m}$) does not seemingly increase much at $z > 2$, being around 0.02\% for a range of redshifts $> 1.4$. The solution to this seeming discrepancy between the fraction of quasar pairs and the excess over expectation codified in the quasar autocorrelation is the dwindling space density of quasars at $z > 2$. As can be seen from our estimates of $\dot{\phi}_\xi$, which, we note, are purely illustrative, the merger rates implied by our best guesses at the relevant parameters hold constant, or even increase slightly from $z > 2$ to $z < 2$. Hence, the changing space density of quasars means that a large amplitude for the quasar autocorrelation at $z > 2$, at separations $\leqsim~35\kpch$, is not inconsistent with a steady fraction of quasar binaries as a function of redshift.

\begin{deluxetable*}{cccccccccccc}
\tabletypesize{\scriptsize}
\tablecaption{The evolution of quasar clustering at $2.4'' < \theta < 6''$ ($\sim$14 to $35\kpch$) with photometric redshift $\zp$\label{table:2}}
\tablecolumns{9}
\tablewidth{0pt}
\tablehead{
 \colhead{$\bzp$} & \colhead{$\omega$\tablenotemark{a}} & \colhead{$A$\tablenotemark{b}} & \colhead{$b_Q$}  & \colhead{$2f_m\tablenotemark{c}/10^{-4}$} & \colhead{$N/n_{Q}$\tablenotemark{d}}  & \colhead{$N_Q$\tablenotemark{e}} &  \colhead{$n_{Q}$\tablenotemark{f}} & \colhead{$n_{\phi}$\tablenotemark{g}} & \colhead{$\dot{\phi}_{f}$\tablenotemark{h}} & \colhead{$\dot{\phi}_{\xi}$\tablenotemark{i}}   }
\startdata
1.40\tablenotemark{j} & $0.58\pm{0.17}$ & $2.71\pm{0.77}$ & $3.96\pm{0.57}$ & $<6.4\pm^{0.7}_{0.7}$ & $1.4\pm0.4$ & 45.2 & 5.7 & & $\leqsim$5.2 & $\sim$0.13 \\
0.75 & $1.60\pm{0.98}$ & $4.07\pm{2.50}$ & $3.89\pm{1.20}$ & $<2.1\pm^{1.1}_{0.8}$ ($2.1\pm^{1.1}_{0.8}$) & $1.9\pm{1.2} $ & 9.9 &$8.8$& 59 &$\leqsim2.6$& $\sim$0.42 \\
1.20 & $4.02\pm{1.67}$ & $5.61\pm{2.34}$ & $4.86\pm{1.01}$ & $<2.9\pm^{1.3}_{0.9}$ ($4.5\pm^{1.6}_{1.2}$) & $2.1\pm{0.9} $ & 9.4 &$7.3$& 14 &$\leqsim$3.0& $\sim$0.32 \\
1.53 & $1.23\pm{1.00}$ & $2.32\pm{1.88}$ & $3.40\pm{1.38}$ & $<1.6\pm^{1.1}_{0.6}$ ($7.6\pm^{1.9}_{1.6}$) & $0.85\pm{0.69}$ & 9.1 &$8.1$& 12 &$\leqsim$1.9& $\sim$0.16  \\
1.87 & $0.73\pm{0.71}$ & $1.03\pm{0.99}$ & $2.85\pm{1.38}$ & $<1.6\pm^{1.0}_{0.6}$ ($4.0\pm^{1.3}_{1.0}$) & $0.49\pm{0.47}$ & 11.0 &$6.9$& 10 &$\leqsim$1.6& $\sim$0.07 \\
2.28 & $6.49\pm{4.32}$ & $13.4\pm{8.9}$ & $10.4\pm{3.5}$ & $<2.1\pm^{2.0}_{1.1}$ ($3.5\pm^{2.4}_{1.5}$) & $6.8\pm{4.6}$ & 4.4 &$1.5$ &  &$\leqsim$0.45& $\sim$0.04  \\
\enddata

\tablenotetext{a}{The angular autocorrelation of quasars integrated over the scale of interest.}
\tablenotetext{b}{The amplitude of small-scale clustering/the amplitude of the best-fit large-scale $b_Q$ model.}
\tablenotetext{c}{(Upper limit on) fraction of UVX quasars that are tracing merging galaxies. Numbers in parentheses attempt to correct for our use of photometric redshifts, and are based on the integrated overlap of primary photometric redshift solutions for a quasar pair in each bin. Errors are Poisson (e.g., \citealt{Geh86}).} 
\tablenotetext{d}{Errors are set purely by $b_Q$. Units are $h^{-3}{\rm Mpc}^{3}$. Note that dividing these values by the volume in a sphere of radius $0.035\Mpch$ ($\sim1/5568$) estimates the integrated real-space correlation function out to $0.035\Mpch$.}
\tablenotetext{e}{Number of quasars in the bin per square degree}
\tablenotetext{f}{Comoving space density of quasars (no correction) in units of $10^{-6}h^{3}{\rm Mpc}^{-3}$}
\tablenotetext{g}{Comoving space density corrected via \citet{Ric05} luminosity function, where valid. Units are $10^{-6}h^{3}{\rm Mpc}^{-3}$.}
\tablenotetext{h}{(Upper limit on) comoving merger rate based on $f_m$ in units of  $h^{4}{\rm Gyr}^{-1}{\rm Gpc}^{3}$. We use $m=3$ ($M_{DMH} = m\times10^{12}~h^{-1}M_\sun$).}
\tablenotetext{i}{Comoving merger rate based on $\xi_{QQ}$ and $n_{Q}$ in units of  $h^{4}{\rm Gyr}^{-1}{\rm Gpc}^{3}$. We use $m=3$ ($M_{DMH} = m\times10^{12}~h^{-1}M_\sun$).}
\tablenotetext{j}{Values for the entire ($A_g <0.21$) DR4 KDE sample. Although a tiny fraction of the sample lie at $z_{phot} < 0.4$, these are likely due to catastrophic $z_{phot}$ estimates, so we integrate over $z \geq 0.4$ to determine $n_{Q}$ and $n_\phi$.}

\end{deluxetable*}

\subsection{Comparison With Real-space Small-scale Clustering Excesses}
\label{sec:ultim}

Of particular interest in Table~\ref{table:2} is $N/n_{Q}$, which arises simply from the correlation function and $d_{Q}$ (see Equation~\ref{eqn:Nfrac}). In the table, this is expressed in $h^{-3}{\rm Mpc}^{3}$ but one can derive the implied number of pairs within $d_{Q}$ compared to the expectation in real space (the integrated correlation function in real space) by dividing $N/n_{Q}$ by the volume of a sphere of radius $d_{Q}$ ($V_{Q} \sim 1 h^{-3}{\rm Mpc}^{3}/5568$). This yields, for instance, an apparent excess in our highest redshift bin of $N/N_{Q}\sim41,200\pm 39,000$. To clarify our notation, and how we specifically calculate these ratios; $N/n_{Q} = 4\pi\int r^2\xi_{QQ}dr$ with $\xi_{QQ}$ calculated as in Equation~\ref{eqn:clar} (c.f. Equation~\ref{eqn:Nfrac}); and $N/N_{Q} = \int r^2\xi_{QQ}{\rm d}r/\int r^2{\rm d}r$.

We can use our $N/n_{Q}$ calculations to compare our measurements to the real space clustering excesses found by Hen06 relative to the $\xi(r)=(r/4.8\Mpch)^{-1.53}$ power-law of PMN04. Weighting our values of $N/n_{Q}$ across the range $1.0 < z < 2.1$, we find $N/n_{Q}=0.84\pm0.35$, implying $N/N_{Q}=4740\pm1970$. Note that if we had instead weighted across $0.4 < z < 2.1$, our values would have increased marginally to $N/N_{Q}=5220\pm1890$. The PMN04 power-law (derived over the range $0.8 < z < 2.1$) implies $N/N_{Q}=3800\pm^{750}_{700}$.  This means that on scales of $<35\kpch$ we find a real-space excess of only $1.2\pm0.6$ times.

It might be argued that we underestimate small-scale excesses, as our statistical power is at $>14\kpch$ but our models integrate from $0\kpch$, and the quasar bias may be yet higher on scales smaller than we can probe. Further, photometric image quality, given the SDSS median seeing of $\sim1.4''$, limits pair separations to $\Delta\theta~\geqsim~3''$ (e.g., Hen06). However, we can explicitly calculate our model and data values over just our $3.8''<\Delta\theta<6''$ (17.7 to $35.4\kpch$) bin, and find that our real-space excess increases to a factor of $1.6\pm0.8$. One might also argue that our photometric redshifts scatter genuine quasar binaries across different redshift bins (i.e., see parenthetical values of $f_m$ in Table~\ref{table:2}).  To quantify this effect, we integrate over overlapping primary photometric redshift solutions in each bin, finding that we may understimate the pair fractions by as much as a factor of 2.6 across $1.0 < z < 2.1$. This assumes that all pairs with overlapping photometric redshift solutions are binary quasars (i.e., no chance projections), and effectively distributes $\sim$85\% of our probable quasar binaries across $\sim$65\% of our sampled volume. Incorporating this correction implies a hard upper limit on any excess of $4.3\pm1.3$ (where the error is reduced to account for the extra quasar pairs) over the $\xi(r)=(r/4.8\Mpch)^{-1.53}$ power-law for PMN04.

If we allow for Poisson error of $\pm2.4$ in the factor of 10 excess claimed by Hen06 at $\leqsim~40\kpch$, based on the 18 quasar pairs they study with separation $\leqsim~40\kpch$, we find our hard upper limit differs by $2.2\sigma$ from their measurement on these scales. Our hard upper limit is, however, well within 1$\sigma$ below the steeper power-law fit quoted by PMN04, $\xi(r)=(r/5.4\Mpch)^{-1.8}$, suggesting that steeper power-laws are allowed by our data, just not the larger excesses quoted by Hen06. We note that there are a number of important caveats to our comparison to Hen06:

\begin{enumerate}

\item Our comparison relies on the evolution of Smi03 models and a constant, average $b_Q$. Dark matter evolution may be more complex than this on kiloparsec scales, and quasar clustering evolution more complex still on scales traced by galaxy mergers. The redshift dependence of $\Delta_{\rm NL}^2(k,z)$ in Equation~\ref{eqn:clar} would then be inaccurate for our purposes.

\item Hen06 average over scales that we cannot fit given the SDSS photometric completeness. If, as they claim, there are extremely large excesses on scales below 3$''$ ($\leqsim~14\kpch$), this could drive their entire claimed excess at $\leqsim~40\kpch$.

\end{enumerate}

\noindent However, given that our measured clustering amplitudes are significantly higher at $z > 2$, we are willing to speculate that the strong excess detected by Hen06 is due to a selection effect, driven perhaps by the design of their analysis to target tracers of the Ly$\alpha$ forest (which biases their sample to $z > 2$).

\section{Summary and Conclusion}

We have used a sample of $\sim$300,000 photometrically classified quasars from SDSS DR4, and, in particular, 91 independent pairs with separations of 2.4$''$--6$''$, to study quasar clustering on small scales. Our main results are:

\renewcommand{\labelenumi}{(\alph{enumi})}

\begin{enumerate}

\item On scales of $<0.1'$ ($\leqsim~35\kpch$), we find that quasars clusters slightly higher ($2.2\sigma$) than our best-fit large-scale model. Relative to the models of Smi03, which include a halo term, the required quasar bias is $b_Q = 4.21\pm0.98$ ($b_Q = 3.93\pm0.71$) on scales of $\sim$18$\kpch$ ($\sim$28$\kpch$).

\item A pure power-law model ($\omega = 0.0469\pm0.0048\theta^{-0.912\pm0.045}$) is in agreement with our data on $<0.1'$ scales. Assuming the projected real-space index of such a model would be $\sim$-2 (i.e., $\xi(r) \sim r^{-2}$), steeper than typical theoretical dark matter density relations (e.g., \citealt{Nav97,Moo98}), steeper halo profiles could explain excess quasar clustering on  these scales.

\item An alternative explanation is that quasars are distributed densely within halos, as compared to dark matter, on small scales. This could arise if UVX binaries are quasars being ignited in both of two merging galaxies.

\item Assuming that UVX quasars are formed in mergers that are traced by UVX binary pairs, we demonstrate that, for a typical host halo mass of $M_{DMH} \sim m\times10^{12}~h^{-1}M_\sun$ \citep{Cro05}, a dynamical merger time is $t_d\sim(610\pm260)m^{-1/2}h^{-1}$Myr, in good agreement with merger simulations (e.g., \citealt{Hop06,Con06}). Under similar assumptions, we derive illustrative comoving volume merger rates for the merging galaxies being traced by UVX quasars.

\item Splitting our sample into photometric redshift bins, we find (on scales of $\sim$28$\kpch$), $b_Q = 3.82\pm0.59$ at $z < 2$ rising to at least 10\% higher than $10.4\pm3.5$ at $z > 2$; a growth in quasar clustering at $z > 2$ of $2.0\sigma$ significance.

\item We demonstrate that a sharp increase in $\omega_{QQ}$ at $z > 2$ could be consistent with a near-constant fraction of UVX quasars being in binaries, given reasonable assumptions about merger rates for the galaxies being traced by UVX quasar binaries. The explanation lies simply in the evolution of the space density of quasars.

\item From $\omega_{QQ}$ and $d_{Q}$, we derive the excess of quasar pairs in real space integrated over $17.7$--$35.4\kpch$. Over $1.0 < z < 2.1$ we find an excess of a factor of only $1.6\pm0.8$ over the power-law advocated by PMN04 (for $0.8 < z < 2.1$). Our hard upper limit on any excess over PMN04 is $4.3\pm1.3$, which differs by $2.2\sigma$ (subject to several caveats, discussed in Section~\ref{sec:ultim}) from the factor of 10 excess found by Hen06 on scales $\leqsim~40\kpch$. We speculate that in biasing their selection to $z > 2$ to target tracers of the Ly$\alpha$ forest, Hen06 inflated their clustering amplitudes relative to PMN04. 

\end{enumerate}

In conclusion, we note that a spectroscopic survey of all 111 pairs listed in Table~\ref{table:pairs} would be very informative. Although only 72 of the 111 pairs have overlapping primary solutions for their photometric redshifts (with a mean probability of populating the primary of $\sim$83\%) quasars at different redshifts separated by $\Delta\theta<0.1'$ have additional scientific uses, such as using the background quasar to trace the foreground quasar in absorption. As $(1+\omega_{QQ})\sim2(1+\omega_{SS})\sim2(1+\omega_{QS})$ at $<0.1'$, and the KDE classification efficiency is close to 95\%, at most $\sim$12 of the 111 DR4 KDE pairs should contain stars. Finally, $\sim$0.9\% of matches between the KDE catalog and the 2QZ are NELGs, meaning that $\sim$2 of the 111 pairs might be a NELG-quasar pair. Pairs containing stars or NELGs will be more likely to be in the set of pairs with non-overlapping photometric redshift solutions.

Binary quasar samples are usually perfunctorily constructed as a corollary to lens surveys, and can therefore be, for instance, originally selected from sets of two images of the same color, or compiled from several surveys with different color, redshift, or magnitude selection. For instance, using DR3, Hen06 targeted only $\sim$20--25 of the ($3'' < \Delta\theta<6''$) quasar pairs that would make the selection criteria of the KDE catalog, compared to the 111 DR4 pairs we list in Table~\ref{table:pairs}. This numerical discrepancy may be due to the strict color criterion insisted on by Hen06 to find lensed pairs (with identical colors) rather than UVX quasar binaries (which can have different colors due to, e.g., scatter in the quasar color-redshift relation). Thus, a spectroscopic survey of our 111 DR4 KDE pairs could represent the first redshift survey of quasar binaries with controlled selection, and could strongly constrain quasar dynamics on scales $\leqsim~35\kpch$, particularly as the line-of-sight velocity difference between binaries can place strong lower limits on the dynamical masses of merging systems (e.g., \citealt{Mor99}). Further, such a survey would place interesting limits on the evolution in the numbers, and dynamical masses, of merging gas-rich galaxies, a key component of models of quasar and galaxy formation.

\acknowledgments

ADM and RJB wish to acknowledge support from NASA through grants NAG5-12578 and NNG06GH15G. GTR was supported in part by a Gordon and Betty Moore Fellowship in Data Intensive Science. RCN acknowledges the EU Marie Curie Excellence Chair for support during this work. DPS acknowledges support through NSF grant AST-0607634. The authors made extensive use of the storage and computing facilities at the National Center for Supercomputing Applications and thank the technical staff for their assistance in enabling this work.

Funding for the SDSS and SDSS-II has been provided by the Alfred P. Sloan Foundation, the Participating Institutions, the National Science Foundation, the U.S. Department of Energy, the National Aeronautics and Space Administration, the Japanese Monbukagakusho, the Max Planck Society, and the Higher Education Funding Council for England. The SDSS Web Site is http://www.sdss.org/.

The SDSS is managed by the Astrophysical Research Consortium for the Participating Institutions. The Participating Institutions are the American Museum of Natural History, Astrophysical Institute Potsdam, University of Basel, Cambridge University, Case Western Reserve University, University of Chicago, Drexel University, Fermilab, the Institute for Advanced Study, the Japan Participation Group, Johns Hopkins University, the Joint Institute for Nuclear Astrophysics, the Kavli Institute for Particle Astrophysics and Cosmology, the Korean Scientist Group, the Chinese Academy of Sciences (LAMOST), Los Alamos National Laboratory, the Max-Planck-Institute for Astronomy (MPIA), the Max-Planck-Institute for Astrophysics (MPA), New Mexico State University, Ohio State University, University of Pittsburgh, University of Portsmouth, Princeton University, the United States Naval Observatory, and the University of Washington.

\begin{deluxetable*}{cccccccc}
\tabletypesize{\scriptsize}
\tablecaption{Quasar pairs in the DR4 KDE catalog with separations $< 0.1'$\label{table:pairs}}
\tablecolumns{8}
\tablewidth{0pt}
\tablehead{
\colhead{$\Delta\theta ('')$} & \colhead{Name} & \colhead{$\alpha$ (J2000)} & \colhead{$\delta$ (J2000)} & \colhead{$u$}  & \colhead{$g$} & \colhead{$z_{phot}$ range}& \colhead{$z_{spec}$}
}
\startdata
4.30 & SDSSJ024037.34-070626.3 & 02 40 37.35 & -07 06 26.3 & 20.59 & 20.51 & 1.45,1.875,1.95 & \\ & SDSSJ024037.11-070623.9 & 02 40 37.11 & -07 06 23.9 & 21.53 & 20.98 & 1.45,1.575,1.80 &  \\ \hline
4.90* & SDSSJ024907.77+003917.1 & 02 49 07.78 & +00 39 17.1 & 19.74 & 19.36 & 2.00,2.175,2.25 &2.164\tablenotemark{b} \\ & SDSSJ024907.86+003912.4 & 02 49 07.87 & +00 39 12.4 & 21.10 & 20.63 & 0.45,0.675,0.85 &star\tablenotemark{b} \\ \hline
5.01 & SDSSJ034134.90-063150.3 & 03 41 34.91 & -06 31 50.4 & 19.70 & 19.67 & 1.40,1.925,2.10 & \\ & SDSSJ034134.78-063145.7 & 03 41 34.79 & -06 31 45.7 & 21.16 & 21.18 & 1.05,1.425,1.55 &  \\ \hline
5.93* & SDSSJ071803.50+402102.6 & 07 18 03.51 & +40 21 02.7 & 20.32 & 20.39 & 1.65,1.925,2.10 & \\ & SDSSJ071803.09+402059.1 & 07 18 03.09 & +40 20 59.2 & 21.24 & 21.07 & 1.60,1.775,2.00 &  \\ \hline
3.06* & SDSSJ073009.55+354151.9 & 07 30 09.56 & +35 41 52.0 & 20.68 & 20.70 & 1.65,1.825,2.00 & \\ & SDSSJ073009.66+354149.2 & 07 30 09.66 & +35 41 49.2 & 21.40 & 20.99 & 0.70,0.875,1.15 &  \\ \hline
4.07 & SDSSJ075339.86+173410.6 & 07 53 39.87 & +17 34 10.7 & 19.79 & 19.61 & 0.85,0.975,1.40 & \\ & SDSSJ075340.08+173408.0 & 07 53 40.08 & +17 34 08.0 & 20.53 & 20.48 & 1.05,1.275,1.40 &  \\ \hline
5.04 & SDSSJ081313.10+541647.0 & 08 13 13.11 & +54 16 47.0 & 17.86 & 17.40 & 0.65,0.775,0.90 & \\ & SDSSJ081312.63+541649.9 & 08 13 12.64 & +54 16 49.9 & 20.68 & 20.25 & 0.65,0.725,0.90 &  \\ \hline
3.20 & SDSSJ081803.82+191703.4 & 08 18 03.83 & +19 17 03.5 & 20.31 & 20.14 & 1.50,1.725,2.00 & \\ & SDSSJ081804.05+191703.8 & 08 18 04.05 & +19 17 03.8 & 21.45 & 21.17 & 1.50,1.775,2.05 &  \\ \hline
4.28 & SDSSJ083258.56+323000.0 & 08 32 58.57 & +32 30 00.1 & 20.01 & 19.59 & 0.40,0.425,0.50 & \\ & SDSSJ083258.34+323003.3 & 08 32 58.35 & +32 30 03.4 & 20.84 & 19.96 & 2.75,2.775,2.80 &  \\ \hline
4.09 & SDSSJ083649.45+484150.0 & 08 36 49.46 & +48 41 50.1 & 19.71 & 19.31 & 0.45,0.675,0.80 &0.657\tablenotemark{b} \\ & SDSSJ083649.55+484154.0 & 08 36 49.55 & +48 41 54.1 & 18.76 & 18.50 & 1.50,1.675,1.95 & 1.710\tablenotemark{a}(1.712)\tablenotemark{b} \\ \hline
3.42 & SDSSJ084257.63+473344.7 & 08 42 57.64 & +47 33 44.7 & 20.51 & 20.45 & 0.85,1.775,2.10 & \\ & SDSSJ084257.37+473342.5 & 08 42 57.38 & +47 33 42.6 & 19.33 & 19.00 & 0.50,0.625,0.70 &  1.552\tablenotemark{a}\\ \hline
4.64 & SDSSJ084624.33+270958.3 & 08 46 24.34 & +27 09 58.4 & 21.02 & 20.66 & 1.90,2.175,2.25 & \\ & SDSSJ084624.50+271002.3 & 08 46 24.51 & +27 10 02.4 & 20.97 & 20.62 & 2.05,2.175,2.35 &  \\ \hline
2.50* & SDSSJ085011.87+093122.0 & 08 50 11.87 & +09 31 22.1 & 21.65 & 21.17 & 0.40,0.675,0.90 & \\ & SDSSJ085011.82+093119.6 & 08 50 11.82 & +09 31 19.7 & 21.50 & 21.15 & 0.65,0.775,0.95 &  \\ \hline
4.28 & SDSSJ085914.77+424123.6 & 08 59 14.77 & +42 41 23.7 & 21.01 & 21.02 & 0.95,1.375,1.65 & \\ & SDSSJ085915.15+424123.5 & 08 59 15.16 & +42 41 23.6 & 19.40 & 19.22 & 0.80,0.975,1.40 & 0.897\tablenotemark{a,?}  \\ \hline
4.68 & SDSSJ090018.12+031228.5 & 09 00 18.13 & +03 12 28.5 & 19.77 & 19.65 & 1.05,1.425,1.50 &1.338\tablenotemark{a} \\ & SDSSJ090017.91+031231.9 & 09 00 17.91 & +03 12 32.0 & 21.02 & 20.79 & 1.40,1.575,2.00 &  \\ \hline
5.42 & SDSSJ090235.73+563756.2 & 09 02 35.73 & +56 37 56.3 & 20.60 & 20.56 & 1.05,1.225,1.45 &1.340\tablenotemark{b} \\ & SDSSJ090235.35+563751.8 & 09 02 35.36 & +56 37 51.8 & 20.75 & 20.95 & 1.15,1.275,1.45 &1.390\tablenotemark{b}  \\ \hline
4.80 & SDSSJ092456.85+092003.0 & 09 24 56.85 & +09 20 03.1 & 20.75 & 20.71 & 1.05,1.375,1.50 & \\ & SDSSJ092457.08+092006.3 & 09 24 57.09 & +09 20 06.4 & 21.41 & 21.05 & 1.45,1.525,1.80 &  \\ \hline
4.04* & SDSSJ092544.30+052541.6 & 09 25 44.30 & +05 25 41.6 & 21.05 & 20.49 & 0.65,0.875,1.00 & \\ & SDSSJ092544.41+052545.3 & 09 25 44.41 & +05 25 45.3 & 20.53 & 20.71 & 1.20,1.425,1.50 &  \\ \hline
4.82* & SDSSJ092659.17+070652.1 & 09 26 59.17 & +07 06 52.1 & 20.56 & 20.40 & 0.95,1.375,1.45 & \\ & SDSSJ092659.28+070656.6 & 09 26 59.29 & +07 06 56.6 & 20.97 & 20.93 & 0.95,1.375,1.50 &  \\ \hline
5.50 & SDSSJ093015.01+420033.6 & 09 30 15.02 & +42 00 33.7 & 19.98 & 20.03 & 1.85,2.025,2.15 & \\ & SDSSJ093014.81+420038.7 & 09 30 14.82 & +42 00 38.7 & 19.81 & 19.71 & 1.15,1.425,1.50 &  \\ \hline
4.80 & SDSSJ093424.32+421130.8 & 09 34 24.32 & +42 11 30.9 & 20.36 & 20.30 & 1.00,1.125,1.40 & \\ & SDSSJ093424.11+421135.0 & 09 34 24.11 & +42 11 35.0 & 21.25 & 21.01 & 1.45,1.675,2.20 &  \\ \hline
5.55 & SDSSJ093521.02+641219.8 & 09 35 21.02 & +64 12 19.9 & 21.02 & 20.99 & 1.45,1.775,2.10 & \\ & SDSSJ093521.80+641221.9 & 09 35 21.81 & +64 12 22.0 & 21.30 & 20.96 & 0.25,0.375,0.45 &  \\ \hline
5.34 & SDSSJ093735.59+631458.6 & 09 37 35.60 & +63 14 58.6 & 21.16 & 20.99 & 0.95,1.325,1.45 & \\ & SDSSJ093734.88+631456.3 & 09 37 34.88 & +63 14 56.3 & 20.35 & 20.22 & 0.85,1.025,1.40 &  \\ \hline
5.59 & SDSSJ093804.87+531742.7 & 09 38 04.87 & +53 17 42.8 & 20.08 & 19.63 & 2.10,2.225,2.45 & \\ & SDSSJ093804.25+531743.6 & 09 38 04.26 & +53 17 43.6 & 20.70 & 20.82 & 1.70,2.025,2.15 &  \\ \hline
4.57 & SDSSJ094309.66+103400.6 & 09 43 09.67 & +10 34 00.6 & 19.44 & 19.24 & 1.05,1.325,1.40 &1.239\tablenotemark{a} \\ & SDSSJ094309.36+103401.3 & 09 43 09.36 & +10 34 01.3 & 20.46 & 20.07 & 0.95,1.525,1.70 &  \\ \hline
3.14 & SDSSJ095454.73+373419.7 & 09 54 54.74 & +37 34 19.8 & 19.81 & 19.57 & 0.95,1.475,1.65 &1.554\tablenotemark{b} \\ & SDSSJ095454.99+373419.9 & 09 54 55.00 & +37 34 20.0 & 19.19 & 18.91 & 1.45,1.575,1.95 &1.884\tablenotemark{a}(1.892)\tablenotemark{b}  \\ \hline
5.33 & SDSSJ095840.74+332216.3 & 09 58 40.75 & +33 22 16.3 & 19.27 & 19.18 & 1.35,1.475,2.10 & \\ & SDSSJ095840.94+332211.5 & 09 58 40.95 & +33 22 11.6 & 20.75 & 20.64 & 1.45,1.725,2.10 &  \\ \hline
3.94 & SDSSJ095907.05+544908.0 & 09 59 07.06 & +54 49 08.1 & 20.51 & 20.60 & 1.90,2.025,2.15 &1.956\tablenotemark{b} \\ & SDSSJ095907.47+544906.3 & 09 59 07.47 & +54 49 06.4 & 20.27 & 20.07 & 1.40,1.575,2.10 &1.954\tablenotemark{b}  \\ \hline
2.93 & SDSSJ100128.61+502756.8 & 10 01 28.61 & +50 27 56.9 & 17.74 & 17.60 & 1.45,1.725,2.00 &1.839\tablenotemark{a,??} \\ & SDSSJ100128.34+502758.4 & 10 01 28.35 & +50 27 58.4 & 18.64 & 18.34 & 1.45,1.575,1.95 &  \\ \hline
3.76 & SDSSJ100434.79+411239.2 & 10 04 34.80 & +41 12 39.3 & 18.78 & 18.64 & 1.55,1.725,2.00 & \\ & SDSSJ100434.91+411242.8 & 10 04 34.92 & +41 12 42.8 & 19.25 & 19.04 & 1.55,1.725,2.15 &1.740\tablenotemark{a,??}  \\ \hline
4.19 & SDSSJ100735.42+012559.3 & 10 07 35.42 & +01 25 59.3 & 21.19 & 21.04 & 0.95,1.225,1.45 & \\ & SDSSJ100735.58+012555.9 & 10 07 35.59 & +01 25 55.9 & 20.49 & 20.58 & 1.00,1.225,1.45 &  \\ \hline
2.93 & SDSSJ103340.52+022831.0 & 10 33 40.52 & +02 28 31.1 & 20.50 & 20.25 & 1.50,1.775,1.95 & \\ & SDSSJ103340.32+022830.6 & 10 33 40.33 & +02 28 30.6 & 20.79 & 20.26 & 0.25,0.375,0.45 &  \\ \hline
3.46 & SDSSJ103939.31+100253.0 & 10 39 39.32 & +10 02 53.0 & 18.88 & 18.42 & 0.10,0.175,0.25 &0.161\tablenotemark{a} \\ & SDSSJ103939.53+100254.3 & 10 39 39.53 & +10 02 54.4 & 19.79 & 19.60 & 0.65,1.125,1.55 &  \\ \hline
5.79 & SDSSJ104110.01+051443.7 & 10 41 10.01 & +05 14 43.7 & 21.34 & 20.61 & 2.20,2.375,2.70 & \\ & SDSSJ104109.71+051447.5 & 10 41 09.72 & +05 14 47.6 & 20.91 & 20.61 & 0.10,0.175,0.25 &  \\ \hline
2.68 & SDSSJ104421.06+042950.8 & 10 44 21.07 & +04 29 50.8 & 21.16 & 20.97 & 0.90,1.575,1.95 & \\ & SDSSJ104421.22+042949.6 & 10 44 21.23 & +04 29 49.7 & 19.78 & 19.86 & 1.10,1.425,1.50 &  \\ \hline
5.37 & SDSSJ104530.70+510611.4 & 10 45 30.70 & +51 06 11.4 & 20.72 & 20.76 & 1.00,1.225,1.45 & \\ & SDSSJ104531.09+510615.3 & 10 45 31.09 & +51 06 15.4 & 20.82 & 20.39 & 1.45,1.575,1.80 &  \\
\hline
\enddata
\end{deluxetable*}

\setcounter{table}{1}

\begin{deluxetable*}{cccccccc}
\tabletypesize{\scriptsize}
\tablecaption{}
\tablecolumns{8}
\tablewidth{0pt}
\tablehead{
\colhead{$\Delta\theta ('')$} & \colhead{Name} & \colhead{$\alpha$ (J2000)} & \colhead{$\delta$ (J2000)} & \colhead{$u$}  & \colhead{$g$} & \colhead{$z_{phot}$ range}& \colhead{$z_{spec}$}
}
\startdata
3.59 & SDSSJ110352.16+443604.2 & 11 03 52.17 & +44 36 04.2 & 20.62 & 20.43 & 0.95,1.225,1.45 & \\ & SDSSJ110351.86+443602.5 & 11 03 51.87 & +44 36 02.6 & 20.88 & 20.62 & 1.40,1.525,1.80 &  \\ \hline
5.91 & SDSSJ111054.10+605343.1 & 11 10 54.11 & +60 53 43.2 & 21.06 & 20.86 & 1.55,1.775,2.20 & \\ & SDSSJ111053.63+605347.9 & 11 10 53.63 & +60 53 48.0 & 19.25 & 18.94 & 0.65,0.775,0.90 &  \\ \hline
5.74 & SDSSJ111844.67+142850.6 & 11 18 44.67 & +14 28 50.7 & 21.20 & 21.01 & 1.40,1.625,2.20 & \\ & SDSSJ111844.40+142854.8 & 11 18 44.40 & +14 28 54.9 & 21.17 & 21.08 & 1.45,1.875,2.10 &  \\ \hline
5.19 & SDSSJ112556.32+143148.0 & 11 25 56.32 & +14 31 48.1 & 20.98 & 20.70 & 0.90,1.325,1.60 & \\ & SDSSJ112556.54+143152.1 & 11 25 56.55 & +14 31 52.1 & 20.49 & 20.38 & 1.50,1.725,2.15 &  \\ \hline
5.78 & SDSSJ113638.09+563503.9 & 11 36 38.09 & +56 35 03.9 & 19.48 & 19.02 & 0.60,0.675,0.90 & \\ & SDSSJ113637.52+563500.4 & 11 36 37.53 & +56 35 00.5 & 19.96 & 19.94 & 1.85,2.075,2.15 &  \\ \hline
4.92 & SDSSJ114503.06+660211.3 & 11 45 03.06 & +66 02 11.3 & 20.02 & 20.09 & 1.65,1.825,2.00 & \\ & SDSSJ114503.74+660208.6 & 11 45 03.74 & +66 02 08.7 & 20.74 & 20.24 & 0.50,0.675,0.85 &  \\ \hline
4.76 & SDSSJ114718.66+123436.3 & 11 47 18.67 & +12 34 36.3 & 20.30 & 19.80 & 2.15,2.225,2.60 &2.232\tablenotemark{b} \\ & SDSSJ114718.44+123439.8 & 11 47 18.45 & +12 34 39.8 & 21.22 & 20.91 & 1.45,1.625,2.00 &1.583\tablenotemark{b}  \\ \hline
1.90 & SDSSJ115119.05+465234.5 & 11 51 19.05 & +46 52 34.5 & 21.48 & 20.50 & 0.30,0.325,0.40 & \\ & SDSSJ115118.87+465235.1 & 11 51 18.88 & +46 52 35.1 & 20.77 & 20.67 & 0.95,1.225,2.10 &  \\ \hline
3.98 & SDSSJ115217.31-004746.2 & 11 52 17.32 & -00 47 46.2 & 21.51 & 21.00 & 1.45,1.525,1.60 & \\ & SDSSJ115217.32-004750.1 & 11 52 17.32 & -00 47 50.2 & 21.17 & 20.79 & 2.00,2.075,2.20 &  \\ \hline
3.56 & SDSSJ115822.77+123518.5 & 11 58 22.78 & +12 35 18.6 & 20.32 & 19.90 & 0.45,0.525,0.65 & \\ & SDSSJ115822.98+123520.3 & 11 58 22.99 & +12 35 20.3 & 20.63 & 20.12 & 0.40,0.475,0.65 &  \\ \hline
3.06 & SDSSJ120450.54+442835.8 & 12 04 50.54 & +44 28 35.9 & 19.15 & 19.04 & 0.95,1.125,1.45 &1.142\tablenotemark{a}(1.144)\tablenotemark{b} \\ & SDSSJ120450.78+442834.2 & 12 04 50.78 & +44 28 34.2 & 19.63 & 19.48 & 1.35,1.725,1.95 &1.814\tablenotemark{b}  \\ \hline
5.85 & SDSSJ120517.34+062125.3 & 12 05 17.35 & +06 21 25.3 & 20.31 & 20.13 & 1.45,1.625,1.90 & \\ & SDSSJ120517.71+062127.1 & 12 05 17.72 & +06 21 27.1 & 21.12 & 20.79 & 0.80,0.975,1.55 &  \\ \hline
3.04 & SDSSJ120629.65+433220.6 & 12 06 29.65 & +43 32 20.6 & 19.70 & 19.38 & 1.95,2.175,2.35 & \\ & SDSSJ120629.64+433217.5 & 12 06 29.65 & +43 32 17.6 & 18.68 & 18.78 & 1.65,1.825,2.05 &1.789\tablenotemark{a,??}  \\ \hline
3.95 & SDSSJ120727.09+140817.1 & 12 07 27.10 & +14 08 17.2 & 20.61 & 20.39 & 1.60,1.775,2.00 & \\ & SDSSJ120727.25+140820.3 & 12 07 27.26 & +14 08 20.4 & 20.47 & 20.27 & 1.55,1.775,1.95 &  \\ \hline
5.86 & SDSSJ120807.68+124602.3 & 12 08 07.68 & +12 46 02.4 & 19.29 & 19.02 & 1.85,2.075,2.20 &1.742\tablenotemark{a} \\ & SDSSJ120808.04+124605.0 & 12 08 08.04 & +12 46 05.0 & 21.32 & 20.91 & 1.45,1.525,1.70 &  \\ \hline
3.83 & SDSSJ120845.14+351051.0 & 12 08 45.14 & +35 10 51.1 & 18.92 & 18.99 & 1.10,1.275,1.45 & \\ & SDSSJ120844.97+351047.8 & 12 08 44.98 & +35 10 47.8 & 21.49 & 21.02 & 1.45,1.575,1.70 &  \\ \hline
2.83 & SDSSJ120957.67+113659.3 & 12 09 57.67 & +11 36 59.4 & 20.95 & 20.70 & 0.80,0.975,1.50 & \\ & SDSSJ120957.71+113656.6 & 12 09 57.72 & +11 36 56.6 & 20.97 & 20.77 & 0.90,1.125,1.50 &  \\ \hline
1.45 & SDSSJ121646.05+352941.5 & 12 16 46.05 & +35 29 41.5 & 19.56 & 19.43 & 1.90,2.075,2.20 & \\ & SDSSJ121645.93+352941.5 & 12 16 45.93 & +35 29 41.6 & 20.49 & 20.39 & 1.05,1.575,1.95 &  \\ \hline
3.26 & SDSSJ123122.37+493430.7 & 12 31 22.38 & +49 34 30.8 & 20.05 & 19.94 & 1.55,1.775,2.15 & \\ & SDSSJ123122.27+493433.8 & 12 31 22.28 & +49 34 33.9 & 21.28 & 20.63 & 0.50,0.725,0.90 &  \\ \hline
3.51 & SDSSJ123555.27+683627.0 & 12 35 55.27 & +68 36 27.1 & 20.16 & 19.70 & 0.50,0.625,1.10 & \\ & SDSSJ123554.78+683624.7 & 12 35 54.78 & +68 36 24.8 & 19.61 & 19.04 & 2.75,2.775,2.80 &  \\ \hline
5.04 & SDSSJ124948.17+060714.0 & 12 49 48.18 & +06 07 14.0 & 20.48 & 20.38 & 1.85,2.075,2.20 &2.376\tablenotemark{b} \\ & SDSSJ124948.12+060709.0 & 12 49 48.13 & +06 07 09.0 & 21.03 & 20.41 & 2.20,2.325,2.65 &2.001\tablenotemark{b}  \\ \hline
3.01 & SDSSJ125530.44+630900.5 & 12 55 30.44 & +63 09 00.5 & 20.55 & 20.30 & 1.50,1.675,1.90 &1.753\tablenotemark{b} \\ & SDSSJ125530.82+630902.0 & 12 55 30.82 & +63 09 02.1 & 20.55 & 20.60 & 1.10,1.375,1.50 &1.393\tablenotemark{b}  \\ \hline
3.55 & SDSSJ125955.46+124151.0 & 12 59 55.46 & +12 41 51.1 & 20.32 & 19.99 & 1.95,2.175,2.30 &2.180\tablenotemark{b} \\ & SDSSJ125955.61+124153.8 & 12 59 55.62 & +12 41 53.8 & 20.44 & 20.09 & 1.90,2.175,2.25 &2.189\tablenotemark{b}  \\ \hline
5.21 & SDSSJ130036.43+082802.9 & 13 00 36.44 & +08 28 02.9 & 17.83 & 17.75 & 1.00,1.125,1.40 & \\ & SDSSJ130036.15+082759.8 & 13 00 36.15 & +08 27 59.9 & 21.17 & 21.07 & 0.75,1.025,1.40 &  \\ \hline
3.81 & SDSSJ130326.14+510051.0 & 13 03 26.14 & +51 00 51.0 & 20.66 & 20.54 & 1.50,2.075,2.20 &1.686\tablenotemark{b} \\ & SDSSJ130326.17+510047.2 & 13 03 26.18 & +51 00 47.2 & 20.47 & 20.37 & 1.60,1.775,2.00 &1.684\tablenotemark{b}  \\ \hline
4.74 & SDSSJ132022.54+305622.8 & 13 20 22.55 & +30 56 22.9 & 18.83 & 18.60 & 1.35,1.475,1.65 & \\ & SDSSJ132022.64+305618.2 & 13 20 22.64 & +30 56 18.3 & 20.33 & 19.92 & 1.45,1.575,1.80 &  \\ \hline
3.06 & SDSSJ133114.29+143834.4 & 13 31 14.29 & +14 38 34.4 & 20.96 & 20.71 & 1.50,2.125,2.20 & \\ & SDSSJ133114.19+143837.0 & 13 31 14.19 & +14 38 37.1 & 21.75 & 21.07 & 0.40,0.475,0.70 &  \\ \hline
5.27 & SDSSJ133128.86+373714.4 & 13 31 28.87 & +37 37 14.4 & 21.65 & 21.02 & 0.35,0.425,0.55 & \\ & SDSSJ133129.01+373709.4 & 13 31 29.02 & +37 37 09.4 & 20.29 & 20.30 & 1.55,1.775,2.15 &  \\ \hline
3.12 & SDSSJ133713.08+601209.6 & 13 37 13.08 & +60 12 09.7 & 20.23 & 20.04 & 1.30,1.775,2.05 &1.721\tablenotemark{b} \\ & SDSSJ133713.13+601206.5 & 13 37 13.13 & +60 12 06.6 & 18.77 & 18.59 & 1.50,1.625,1.95 &1.735\tablenotemark{a}(1.727)\tablenotemark{b} \\ \hline
3.89 & SDSSJ133901.97+620851.5 & 13 39 01.98 & +62 08 51.6 & 20.55 & 20.32 & 1.50,1.775,1.90 & \\ & SDSSJ133901.91+620847.7 & 13 39 01.91 & +62 08 47.7 & 21.27 & 20.89 & 1.45,1.575,1.95 &  \\ \hline
1.69 & SDSSJ133907.13+131039.6 & 13 39 07.14 & +13 10 39.6 & 19.13 & 18.85 & 0.65,0.875,1.05 & \\ & SDSSJ133907.23+131038.6 & 13 39 07.23 & +13 10 38.7 & 19.62 & 19.03 & 2.10,2.325,2.60 &  \\ \hline
2.99 & SDSSJ134929.84+122706.9 & 13 49 29.85 & +12 27 06.9 & 18.01 & 17.87 & 1.50,1.725,1.95 &1.722\tablenotemark{a,b} \\ & SDSSJ134930.00+122708.8 & 13 49 30.01 & +12 27 08.8 & 19.73 & 19.33 & 1.45,1.575,2.20 &1.722\tablenotemark{b}  \\ \hline
4.50 & SDSSJ141855.41+244108.9 & 14 18 55.42 & +24 41 08.9 & 19.81 & 19.27 & 0.45,0.525,0.70 & \\ & SDSSJ141855.53+244104.7 & 14 18 55.54 & +24 41 04.7 & 20.84 & 20.22 & 0.40,0.625,0.70 &  \\ \hline
4.94 & SDSSJ142359.48+545250.8 & 14 23 59.48 & +54 52 50.8 & 18.92 & 18.63 & 1.00,1.175,1.45 &1.409\tablenotemark{a} \\ & SDSSJ142400.00+545248.7 & 14 24 00.01 & +54 52 48.8 & 20.29 & 19.93 & 1.45,1.575,1.90 &  \\ \hline
4.27 & SDSSJ142604.32+071930.0 & 14 26 04.33 & +07 19 30.0 & 20.09 & 20.12 & 1.00,1.225,1.45 & \\ & SDSSJ142604.26+071925.8 & 14 26 04.27 & +07 19 25.9 & 20.77 & 20.82 & 0.95,1.175,1.45 &  \\ \hline
5.41 & SDSSJ143002.88+071411.3 & 14 30 02.89 & +07 14 11.3 & 19.56 & 19.50 & 1.05,1.375,1.45 & \\ & SDSSJ143002.66+071415.6 & 14 30 02.66 & +07 14 15.6 & 20.58 & 20.27 & 1.00,1.225,1.40 &  \\ \hline
5.91 & SDSSJ143104.97+270528.6 & 14 31 04.98 & +27 05 28.6 & 20.91 & 20.24 & 2.20,2.375,2.65 & \\ & SDSSJ143104.64+270524.6 & 14 31 04.65 & +27 05 24.7 & 20.55 & 19.79 & 2.25,2.375,2.55 &  \\ 
\hline
\enddata
\end{deluxetable*}

\setcounter{table}{1}

\newpage

\begin{deluxetable*}{cccccccc}
\tabletypesize{\scriptsize}
\tablecaption{}
\tablecolumns{8}
\tablewidth{0pt}
\tablehead{
\colhead{$\Delta\theta ('')$} & \colhead{Name} & \colhead{$\alpha$ (J2000)} & \colhead{$\delta$ (J2000)} & \colhead{$u$}  & \colhead{$g$} & \colhead{$z_{phot}$ range}& \colhead{$z_{spec}$}
}
\startdata
5.13 & SDSSJ143229.24-010616.0 & 14 32 29.25 & -01 06 16.1 & 17.88 & 17.83 & 1.90,2.025,2.15 &2.085\tablenotemark{a}(2.082)\tablenotemark{b} \\ & SDSSJ143228.94-010613.5 & 14 32 28.95 & -01 06 13.6 & 21.40 & 21.10 & 1.55,2.125,2.25 &2.082\tablenotemark{b}  \\ \hline
5.33 & SDSSJ143949.67+060107.9 & 14 39 49.67 & +06 01 08.0 & 19.70 & 19.56 & 0.90,1.025,1.40 & \\ & SDSSJ143949.83+060103.2 & 14 39 49.83 & +06 01 03.2 & 21.14 & 20.94 & 1.00,1.375,1.50 &  \\ \hline
3.45 & SDSSJ144422.59+541320.6 & 14 44 22.59 & +54 13 20.6 & 20.53 & 20.31 & 0.70,0.925,1.15 & \\ & SDSSJ144422.21+541321.4 & 14 44 22.21 & +54 13 21.5 & 21.18 & 20.83 & 1.45,1.575,1.80 &  \\ \hline
3.99 & SDSSJ144740.62+632732.9 & 14 47 40.62 & +63 27 33.0 & 21.03 & 21.00 & 0.90,1.225,1.50 & \\ & SDSSJ144741.09+632735.3 & 14 47 41.10 & +63 27 35.4 & 19.54 & 19.52 & 1.05,1.225,1.45 &  \\ \hline
5.14 & SDSSJ145826.72+544813.1 & 14 58 26.73 & +54 48 13.2 & 20.82 & 20.53 & 1.65,1.925,1.95 & \\ & SDSSJ145826.16+544814.8 & 14 58 26.17 & +54 48 14.8 & 21.08 & 20.79 & 1.50,1.775,1.95 &  \\ \hline
4.00 & SDSSJ150656.86+505610.5 & 15 06 56.87 & +50 56 10.6 & 19.47 & 19.22 & 0.65,0.775,1.00 & \\ & SDSSJ150657.18+505607.9 & 15 06 57.18 & +50 56 07.9 & 20.18 & 19.75 & 2.00,2.225,2.40 &  \\ \hline
4.35 & SDSSJ150747.23+290333.2 & 15 07 47.23 & +29 03 33.3 & 20.40 & 19.97 & 0.70,0.775,0.95 & \\ & SDSSJ150746.90+290334.1 & 15 07 46.91 & +29 03 34.2 & 20.70 & 20.44 & 0.80,0.975,1.25 &  \\ \hline
2.90 & SDSSJ150842.21+332805.5 & 15 08 42.22 & +33 28 05.5 & 20.70 & 20.43 & 0.25,0.375,0.45 &0.878\tablenotemark{b} \\ & SDSSJ150842.19+332802.6 & 15 08 42.20 & +33 28 02.6 & 17.95 & 17.83 & 0.70,0.875,1.00 &0.878\tablenotemark{a,b}  \\ \hline
5.62 & SDSSJ151055.97+374124.1 & 15 10 55.97 & +37 41 24.2 & 19.88 & 19.64 & 0.90,1.075,1.45 & \\ & SDSSJ151055.73+374119.3 & 15 10 55.73 & +37 41 19.3 & 20.13 & 19.85 & 0.80,0.975,1.40 &  \\ \hline
5.31 & SDSSJ151258.43+295150.4 & 15 12 58.43 & +29 51 50.4 & 18.75 & 18.68 & 1.45,1.725,2.00 & \\ & SDSSJ151258.06+295148.1 & 15 12 58.07 & +29 51 48.1 & 21.23 & 20.82 & 0.45,0.475,0.50 &  \\ \hline
5.39 & SDSSJ151709.88+084414.3 & 15 17 09.89 & +08 44 14.3 & 21.55 & 20.89 & 0.40,0.475,0.55 & \\ & SDSSJ151709.55+084416.5 & 15 17 09.56 & +08 44 16.5 & 20.94 & 20.91 & 1.70,1.875,1.95 &  \\ \hline
5.28 & SDSSJ151823.05+295925.4 & 15 18 23.06 & +29 59 25.5 & 19.95 & 19.51 & 1.00,1.175,1.45 &1.249\tablenotemark{a} \\ & SDSSJ151823.43+295927.5 & 15 18 23.43 & +29 59 27.6 & 20.12 & 20.24 & 1.10,1.225,1.45 &  \\ \hline
3.52 & SDSSJ152019.86+234107.1 & 15 20 19.86 & +23 41 07.2 & 20.61 & 20.11 & 0.40,0.475,0.70 & \\ & SDSSJ152020.11+234106.5 & 15 20 20.12 & +23 41 06.6 & 20.33 & 19.79 & 0.65,0.775,0.90 &  \\ \hline
1.94 & SDSSJ152050.04+263740.9 & 15 20 50.04 & +26 37 40.9 & 19.12 & 19.19 & 1.85,2.075,2.15 & \\ & SDSSJ152050.18+263740.8 & 15 20 50.19 & +26 37 40.9 & 19.45 & 19.34 & 1.00,1.425,1.75 &  \\ \hline
4.11 & SDSSJ153038.56+530404.0 & 15 30 38.56 & +53 04 04.0 & 20.94 & 20.56 & 1.45,1.575,1.95 &1.531\tablenotemark{b} \\ & SDSSJ153038.82+530400.6 & 15 30 38.82 & +53 04 00.6 & 20.76 & 20.70 & 1.40,1.725,2.15 &1.533\tablenotemark{b}  \\ \hline
3.60 & SDSSJ154334.27+264657.7 & 15 43 34.28 & +26 46 57.8 & 20.94 & 20.84 & 0.80,0.975,1.45 & \\ & SDSSJ154334.28+264654.1 & 15 43 34.29 & +26 46 54.2 & 20.65 & 20.63 & 1.30,1.925,2.10 &  \\ \hline
3.74 & SDSSJ154515.98+275601.0 & 15 45 15.98 & +27 56 01.1 & 20.13 & 19.84 & 1.45,1.525,1.85 & \\ & SDSSJ154515.73+275559.2 & 15 45 15.73 & +27 55 59.3 & 21.35 & 20.69 & 0.45,0.625,0.70 &  \\ \hline
5.37 & SDSSJ155908.39+264031.8 & 15 59 08.39 & +26 40 31.9 & 20.45 & 20.02 & 0.75,0.925,1.10 & \\ & SDSSJ155908.21+264036.7 & 15 59 08.22 & +26 40 36.7 & 21.11 & 20.44 & 2.20,2.275,2.65 &  \\ \hline
3.58 & SDSSJ160032.31+163347.2 & 16 00 32.31 & +16 33 47.2 & 21.00 & 20.79 & 1.45,1.725,2.00 & \\ & SDSSJ160032.54+163348.5 & 16 00 32.54 & +16 33 48.5 & 21.51 & 21.11 & 1.45,1.575,1.90 &  \\ \hline
3.45 & SDSSJ160603.02+290050.8 & 16 06 03.02 & +29 00 50.9 & 18.66 & 18.42 & 0.70,0.875,1.00 & \\ & SDSSJ160602.81+290048.7 & 16 06 02.81 & +29 00 48.8 & 18.87 & 18.50 & 0.50,0.725,1.00 &  \\ \hline
4.57 & SDSSJ160926.28+075324.4 & 16 09 26.29 & +07 53 24.4 & 20.40 & 19.79 & 2.20,2.325,2.60 & \\ & SDSSJ160926.58+075323.1 & 16 09 26.58 & +07 53 23.2 & 21.47 & 21.13 & 0.35,0.725,0.90 &  \\ \hline
5.63 & SDSSJ162419.98+350644.4 & 16 24 19.99 & +35 06 44.5 & 20.83 & 20.72 & 0.75,0.925,1.10 & \\ & SDSSJ162419.80+350649.6 & 16 24 19.81 & +35 06 49.7 & 21.27 & 20.80 & 1.45,1.575,1.90 &  \\ \hline
4.22 & SDSSJ162847.75+413045.4 & 16 28 47.75 & +41 30 45.4 & 20.09 & 19.81 & 1.35,1.525,1.70 & \\ & SDSSJ162848.06+413043.1 & 16 28 48.07 & +41 30 43.2 & 20.60 & 20.40 & 1.95,2.075,2.20 &  \\ \hline
4.35 & SDSSJ162902.59+372430.8 & 16 29 02.59 & +37 24 30.8 & 19.36 & 19.17 & 0.80,0.975,1.10 &0.926\tablenotemark{a}(0.923)\tablenotemark{b} \\ & SDSSJ162902.63+372435.1 & 16 29 02.63 & +37 24 35.2 & 19.53 & 19.35 & 0.70,0.925,1.10 &0.906\tablenotemark{b}  \\ \hline
4.92 & SDSSJ163510.30+291116.1 & 16 35 10.31 & +29 11 16.2 & 20.74 & 20.43 & 1.40,1.525,1.85 & \\ & SDSSJ163510.14+291120.6 & 16 35 10.15 & +29 11 20.6 & 19.14 & 18.83 & 1.45,1.575,1.80 &  \\ \hline
3.90* & SDSSJ163700.93+263609.8 & 16 37 00.93 & +26 36 09.9 & 19.69 & 19.36 & 1.40,1.525,1.80 &1.961\tablenotemark{c} \\ & SDSSJ163700.88+263613.7 & 16 37 00.88 & +26 36 13.7 & 20.99 & 20.61 & 0.45,0.575,0.85 & 1.961\tablenotemark{c}  \\ \hline
5.50 & SDSSJ164130.81+230837.6 & 16 41 30.81 & +23 08 37.7 & 20.36 & 20.35 & 0.85,1.025,1.40 & \\ & SDSSJ164130.90+230843.0 & 16 41 30.91 & +23 08 43.0 & 21.61 & 21.16 & 1.05,1.475,1.60 &  \\ \hline
2.32 & SDSSJ164311.34+315618.3 & 16 43 11.34 & +31 56 18.4 & 19.61 & 19.20 & 1.50,1.675,1.85 &0.586\tablenotemark{a,d} \\ & SDSSJ164311.38+315620.6 & 16 43 11.39 & +31 56 20.6 & 20.45 & 19.99 & 1.35,1.575,1.80 &  \\ \hline
3.62* & SDSSJ164928.79+173306.5 & 16 49 28.79 & +17 33 06.5 & 19.56 & 19.34 & 1.90,2.075,2.20 & \\ & SDSSJ164928.99+173308.5 & 16 49 29.00 & +17 33 08.6 & 20.00 & 19.77 & 1.90,2.075,2.20 &  \\ \hline
5.96* & SDSSJ170735.76+274233.9 & 17 07 35.77 & +27 42 33.9 & 20.46 & 20.03 & 0.40,0.475,0.65 & \\ & SDSSJ170736.04+274238.5 & 17 07 36.05 & +27 42 38.6 & 21.66 & 21.33 & 0.90,1.375,1.50 &  \\ \hline
5.83 & SDSSJ171334.41+553050.3 & 17 13 34.41 & +55 30 50.4 & 19.02 & 18.88 & 1.00,1.375,1.45 &1.277\tablenotemark{a} \\ & SDSSJ171335.03+553047.9 & 17 13 35.04 & +55 30 47.9 & 19.55 & 19.11 & 2.00,2.175,2.20 &  \\ \hline
3.72 & SDSSJ172317.42+590446.4 & 17 23 17.42 & +59 04 46.4 & 19.09 & 18.88 & 1.55,1.725,1.90 &1.600\tablenotemark{a}(1.604)\tablenotemark{b} \\ & SDSSJ172317.30+590442.7 & 17 23 17.31 & +59 04 42.8 & 20.81 & 20.31 & 1.45,1.725,2.25 &1.597\tablenotemark{b}  \\ \hline
0.98* & SDSSJ203718.30-051233.7 & 20 37 18.31 & -05 12 33.7 & 20.40 & 19.72 & 2.80,2.825,4.35 &star\tablenotemark{a} \\ & SDSSJ203718.24-051233.3 & 20 37 18.25 & -05 12 33.3 & 19.51 & 18.52 & 2.35,2.375,4.35 &  \\ \hline
4.95* & SDSSJ205822.47-002003.7 & 20 58 22.47 & -00 20 03.7 & 21.88 & 21.37 & 0.90,1.375,1.55 & \\ & SDSSJ205822.18-002001.3 & 20 58 22.18 & -00 20 01.4 & 21.28 & 21.10 & 1.45,1.725,2.10 &  \\ \hline
5.06* & SDSSJ211157.26+091554.2 & 21 11 57.27 & +09 15 54.3 & 20.41 & 19.83 & 1.00,1.325,1.35 &star\tablenotemark{a} \\ & SDSSJ211157.24+091559.3 & 21 11 57.25 & +09 15 59.3 & 20.64 & 20.73 & 0.95,1.275,1.40 &  \\ \hline
5.81* & SDSSJ221426.79+132652.3 & 22 14 26.79 & +13 26 52.4 & 20.94 & 20.64 & 1.55,2.025,2.20 &1.995\tablenotemark{b} \\ & SDSSJ221427.03+132657.0 & 22 14 27.03 & +13 26 57.0 & 20.48 & 20.34 & 1.65,1.825,2.05 &2.002\tablenotemark{b}  \\ \hline
5.81 & SDSSJ222423.36-094645.4 & 22 24 23.36 & -09 46 45.5 & 20.59 & 20.31 & 1.50,1.675,2.10 & \\ & SDSSJ222423.70-094642.6 & 22 24 23.70 & -09 46 42.6 & 21.16 & 21.04 & 1.45,2.075,2.20 &\\
\hline
\enddata
\end{deluxetable*}

\setcounter{table}{1}

\newpage

\begin{deluxetable*}{cccccccc}
\tabletypesize{\scriptsize}
\tablecaption{}
\tablecolumns{8}
\tablewidth{0pt}
\tablehead{
\colhead{$\Delta\theta ('')$} & \colhead{Name} & \colhead{$\alpha$ (J2000)} & \colhead{$\delta$ (J2000)} & \colhead{$u$}  & \colhead{$g$} & \colhead{$z_{phot}$ range}& \colhead{$z_{spec}$}
}
\startdata
3.52* & SDSSJ230946.52+152145.5 & 23 09 46.53 & +15 21 45.5 & 20.72 & 20.60 & 1.00,1.275,1.45 & \\ & SDSSJ230946.69+152142.9 & 23 09 46.69 & +15 21 43.0 & 21.08 & 20.83 & 1.05,1.325,1.45 &  \\ \hline
\enddata

\tablenotetext{~}{Note that $u$ and $g$ are ``as observed'' (they have {\em not} been corrected for Galactic extinction)}
\tablenotetext{a}{Source: DR4 Catalog Archive Server (e.g., likely \citealt{Sch05}) }
\tablenotetext{b}{Source: \citet{Hen06} } \tablenotetext{c}{Source: \citet{Sra78,Djo84} }
\tablenotetext{d}{Probable binary quasar-starburst \citep{Bro99} }
\tablenotetext{*}{Objects in Galactic regions with absorption $A_g > 0.21$. These were not included in our main analysis, as it is known  (see, e.g., \citealt{Mye06}, or the Appendixes of Paper1) that such objects have a higher probability of being stars}
\tablenotetext{?}{Object with low confidence redshift}
\tablenotetext{??}{High probability lens (as recorded at http://cfa-www.harvard.edu/castles/)}

\end{deluxetable*}


\begin{thebibliography}{}

\bibitem[Abazajian et al.(2003)]{Aba03} Abazajian,~K., et al. 2003, \aj, 126, 2081 (DR1)
\bibitem[Abazajian et al.(2004)]{Aba04} Abazajian,~K., et al. 2004, \aj, 128, 502 (DR2)
\bibitem[Abazajian et al.(2005)]{Aba05} Abazajian,~K., et al. 2005, \aj, 129, 1755 (DR3)
\bibitem[Brotherton et al.(1999)]{Bro99} Brotherton,~M.~S., et al. 1999, \apj, 520, 87
\bibitem[Conselice(2006)]{Con06} Conselice, C.~J. 2006, \apj, 638, 686
\bibitem[Croom et al.(2004)]{Cro04} Croom,~S.~M., Smith,~R.~J., Boyle,~B.~J., Shanks,~T., Miller,~L., Outram,~P.~J., \& Loaring,~N.~S. Ê2004, \mnras, 349, 1397 (2QZ)
\bibitem[Croom et al.(2005)]{Cro05} Croom,~S.~M., et al. 2005, \mnras, 356, 415
\bibitem[Djorgovski(1991)]{Djo91} Djorgovski,~S. 1991, in ASP Conf. Ser. 21, The Space Distribution of Quasars, ed. D. Crampton (San Francisco: ASP), 349
\bibitem[Djorgovski \& Spinrad(1984)]{Djo84} Djorgovski,~S., \& Spinrad,~H. 1984, \apj, 282, L1 
\bibitem[Fukugita et al.(1996)]{Fuk96} Fukugita,~M.,~Ichikawa,~T., Gunn,~J.~E., Doi,~M., Shimasaku,~K., \& Schneider,~D.~P. 1996, \aj, 111, 1748
\bibitem[Gehrels(1986)]{Geh86} Gehrels,~N. 1986, \apj, 303, 336
\bibitem[Gunn et al.(1998)]{Gun98} Gunn,~J.~E., et al. 1998, \aj, 116, 3040
\bibitem[Gunn et al.(2006)]{Gun06} Gunn,~J.~E., et al. 2006, \aj, 131, 2332
\bibitem[Heckman et al.(1986)]{Hec86} Heckman,~T.~M., Smith,~E.~P., Baum,~S.~A., van~Breugel,~W.~J.~M., Miley,~G.~K., Illingworth,~G.~D., Bothun,~G.~D., \& Balick, B. 1986, \apj, 311, 526
\bibitem[Hennawi et al.(2006)]{Hen06} Hennawi,~J.~F., et al. 2006, \aj, 131, 1 (Hen06)
\bibitem[Hogg et al.(2001)]{Hog01} Hogg,~D.~W., Finkbeiner,~D.~P., Schlegel,~D.~J., \& Gunn,~J.~E. 2001, \aj, 122, 2129 
\bibitem[Hopkins et al.(2006)]{Hop06} Hopkins,~P.~F, Hernquist,~L., Cox,~T.~J., Di~Matteo,~T., Robertson,~B.,  \& Springel~V. 2006, \apjs, 163, 1
\bibitem[Ivezic et al.(2004)]{Ive04} Ivezic,~Z.,~et~al. 2004, AN, 325, 583
\bibitem[Kochanek, Falco \& Munez(1999)]{Koc99} Kochanek,~C.~S., Falco,~E.~E., \& Mu\~{n}oz,~J.~A. 1999, \apj, 510, 590
\bibitem[Kormendy \& Richstone(1995)]{Kor95} Kormendy,~J., \& Richstone,~D. 1995, \araa, 33, 581
\bibitem[Lupton, Gunn \& Szalay(1999)]{Lup99} Lupton,~R.~H., Gunn,~J.~E., \& Szalay,~A.~S. 1999, \aj, 118, 1406 
\bibitem[Magorrian et al.(1998)]{Mag98} Magorrian, J., et al. 1998, \aj, 115, 2285
\bibitem[Masjedi et al.(2006)]{Mas06} Masjedi,~M., et al. 2006 \apj, 644, 54
\bibitem[Moore et al.(1998)]{Moo98} Moore,~B., Governato,~F., Quinn,~T., Stadel,~J., \& Lake,~G. 1998, \apj, 499, 5
\bibitem[Mortlock, Webster, \& Francis(1999)]{Mor99} Mortlock,~D.~J., Webster,~R.~L., \& Francis,~P.~J. 1999 \mnras, 309, 836
\bibitem[Myers et al.(2003)]{Mye03} Myers,~A.~D., Outram,~P.~J., Shanks,~T., Boyle,~B.~J., Croom,~S.~M., Loaring,~N.~S., Miller,~L., \& Smith,~R.~J. 2003, \mnras, 342, 467
\bibitem[Myers et al.(2005)]{Mye05} Myers,~A.~D., Outram,~P.~J., Shanks,~T., Boyle,~B.~J., Croom,~S.~M., Loaring,~N.~S., Miller,~L., \& Smith,~R.~J. 2005, \mnras, 359, 741 
\bibitem[Myers et al.(2006)]{Mye06} Myers,~A. D.,~et al. 2006, \apj, 638, 622
\bibitem[Myers et al.(2007)]{Mye07} Myers,~A. D.,~et al. 2007, in preparation (Paper1)
\bibitem[Navarro, Frenk, \& White(1997)]{Nav97} Navarro,~J.~F., Frenk,~C.~S., \& White,~S.~D.~M. 1997, \apj, 490, 493
\bibitem[Peacock \& Smith(2000)]{Pea00} Peacock,~J.~A., \& Smith,~R.~E. 2000, MNRAS, 318, 1144
\bibitem[Pier et al.(2003)]{Pie03} Pier,~J.~R., Munn,~J.~A., Hindsley,~R.~B., Hennessy,~G.~S., Kent,~S.~M., Lupton,~R.~H., \& Ivezic,~Z. 2003, \aj, 125, 1559
\bibitem[Porciani, Magliocchetti \& Norberg(2004)]{Por04} Porciani,~C., Magliocchetti,~M., \& Norberg,~P. 2004, \mnras, 355, 1010 (PMN04)
\bibitem[Richards et al.(2004)]{Ric04} Richards,~G.~T., et al. 2004, \apjs, 155, 257
\bibitem[Richards et al.(2005)]{Ric05} Richards,~G.~T., et al. 2005, \mnras, 360, 839 
\bibitem[Richstone et al.(1998)]{Ric98} Richstone,~D., et al. 1998, \nat, 395, 14
\bibitem[Schlegel, Finkbeiner \& Davis(1998)]{Sch98} Schlegel,~D.~J., Finkbeiner,~D.~P., \& Davis,~M. 1998, \apj, 500, 525
\bibitem[Schneider et al.(2005)]{Sch05} Schneider,~D.~P., et al. 2005, \aj, 130, 367
\bibitem[Schneider(1993)]{Sch93} Schneider,~P. 1993, in Proc. 31st Li\`{e}ge Int. Astroph. Coll., Gravitational Lenses in the Universe, eds. Surdej~J., Fraipont-Caro~D., Gosset~E., Refsdal S., Remy M., 41
\bibitem[Seljak(2000)]{Sel00} Seljak,~U. 2000, \mnras, 318, 203
\bibitem[Silk \& Rees(1998)]{Sil98} Silk,~J., \& Rees,~M.~J. 1998, \aap, 331, 1
\bibitem[Smith et al.(2002)]{Smi02} Smith,~J.~A., et al. 2002, \aj, 123, 2121
\bibitem[Smith et al.(2003)]{Smi03} Smith,~R.~E., et al. 2003, \mnras, 341, 1311 (Smi03)
\bibitem[Sramek \& Weedman(1978)]{Sra78} Sramek,~R.~A., \& Weedman,~D.~W. 1978, \apj, 221, 468
\bibitem[Stoughton et al.(2002)]{Sto02} Stoughton,~C.,~et~al. 2002, \aj, 123, 485
\bibitem[Tucker et al.(2006)]{Tuc06} Tucker,~D., et al. 2006, AN, in press
\bibitem[York et al.(2000)]{Yor00} York,~D.~G., et al. 2000, \aj, 120, 1579

\end{thebibliography}
\end{document}